# Extrinsic suppression of anomalous Hall effect in Fe-rich kagome magnet Fe$_3$Sn

Muhua Liu, Li Ma*, Guoke Li, Congmian Zhen, Denglu Hou, and Dewei Zhao

*Hebei Key Laboratory of Photophysics Research and Application, College of Physics, Hebei Normal University, Shijiazhuang 050024, China*

**Abstract:**

In Fe-based kagome magnets, Fe$_3$Sn has been predicted to have the largest intrinsic anomalous Hall conductivity (AHC) and the highest Curie temperature $T_C$ = 743 K. However, the current experimental results show that the total AHC is much lower than the predicted value due to the strong extrinsic contribution. To suppress the extrinsic contribution and thus enhance the intrinsic contribution, we increased the Fe content in the stoichiometric Fe$_3$Sn. We found that the extrinsic contribution is greatly suppressed, and the intrinsic contribution is dominant and is close to the theoretically predicted value of 555 S/cm over the whole temperature range. Based on the formula of the skew scattering, we analyzed the reason why the skew scattering is suppressed and found that the spin-orbit coupling strength provided by the impurity center is the key. The spin-orbit coupling strength provided by Fe as an impurity center in this study is much smaller than that of Sn as an impurity center in previous studies. Therefore, the AHC in Fe-rich Fe$_3$Sn obeys the unified theory, while the AHC in Sn-rich Fe$_3$Sn deviates from the unified theory. Our study provides a promising solution for the regulation of the extrinsic contribution to the anomalous Hall effect in kagome magnets.

## I. INTRODUCTION

The anomalous Hall effect (AHE) has not only promoted the development of fundamental physics [1,2] but also accelerated the application of materials science and engineering technology [3,4]. Therefore, AHE has long been a research hotspot in physics and materials science. The AHE arises from two contributions: an extrinsic contribution related to asymmetric scattering by impurities, including skew-scattering and side-jump mechanisms, and an intrinsic contribution determined by the Berry phase originating solely from the electronic band structure [5-9]. For metallic systems, if the temperature is low, the relaxation time $\tau$ can be considerable, causing skew scattering to dominate. Conversely, at high temperatures, the skew scattering mechanism is significantly suppressed due to increased phonon scattering [2]. Meanwhile, the intrinsic mechanism, which is independent of scattering, remains unaffected and becomes the primary contributor to AHE [1,2,10,11]. As a result, the intrinsic contribution is robust and provides a reliable solution for the stable application of AHE across various temperatures.

Given that the intrinsic mechanism is governed by the topological band structure of electrons, which in turn is determined by the symmetry of the magnetic and crystal structures, optimizing the magnetic and crystal structure is crucial for addressing the problem. Researchers have concluded that materials with magnetic kagome lattices (kagome magnet) exhibit notable topological energy bands, leading to significant AHE [12]. Experimentally, with the assistance of non-trivial topological bands, limited anomalous Hall conductivity (AHC) has been observed in Mn-based kagome magnets near room temperature, including 20 S/cm for $Mn_3Sn$ [13], 60 S/cm for $Mn_3Ge$ [14], 380 S/cm for $LiMn_6Sn_6$ [15] and 223 S/cm for $GdMn_6Sn_6$ [16]. In the Co-based kagome ferromagnet $Co_3Sn_2S_2$ [17], the value of AHC is as high as 1310 S/cm below the Curie temperature $T_C$ = 177 K. In the Fe-based kagome magnets, the intrinsic anomalous Hall conductivity $\sigma_{xy}^{in}$ of $Fe_3Sn_2$ [18,19] and $Fe_3Ge$ [20] has been found to be 200 S/cm and 449 S/cm, respectively. Recently, different research groups predicted that $\sigma_{xy}^{in}$ in $Fe_3Sn$ is as high as 757 S/cm [21] and 702 S/cm [22], respectively, and the theoretical value in polycrystals is 555 S/cm [23], which is the highest in the entire Fe-Sn family and Fe-based kagome magnets. Additionally, $Fe_3Sn$ has a Curie temperature much higher than room temperature ($T_C$ = 743 K) [24]. However, the maximum AHC experimentally measured in $Fe_3Sn$ single crystals to date is only 500 S/cm, and this value sharply drops to ~300 S/cm at low temperatures

[22,25], which is only about 40% of the predicted value for $Fe_3Sn$ [21,22].

Ref.[25] points out that the low-temperature decrease in AHC is due to the significant extrinsic contribution, which has the opposite sign to the intrinsic contribution. In this context, AHC can be restored by suppressing the extrinsic contribution, which is strongly dependent on impurity scattering centers. It is noteworthy that systems exhibiting a strong extrinsic contribution to AHC are typically Sn-rich ($Fe_{74.9}Sn_{25.1}$ for Ref. [22], $Fe_{74.25}Sn_{25.75}$ for Ref.[23] ). In these cases, Sn atoms occupying normal Fe sites provide the impurity scattering centers. Theoretically, a stoichiometric ratio in $Fe_3Sn$ could effectively eliminate the impurity centers. However, $Fe_3Sn$ belongs to a typical metastable phase [26-28], making it difficult for stoichiometric $Fe_3Sn$ to exist stably; both Fe-rich [27-29] and Sn-rich [27] compositions contribute to stabilizing the hexagonal $Fe_3Sn$ phase.

In this study, we experimentally investigate the charge transport properties of Fe-rich $Fe_3Sn$ and compare our results with those reported in Refs.[22,23,25]. We find that the extrinsic contribution of AHE in the kagome magnet $Fe_3Sn$ is tunable. This work is arranged as follows. Section II describes the experimental methods used in our study. Section III presents the experimental results, focusing on why the anomalous Hall effect is dominated by intrinsic contributions over the entire temperature range. Finally, Section IV gives concluding remarks.

## II. EXPERIMENTAL DETAILS

Polycrystalline $Fe_{3\pm x}Sn$ ($x$ = -0.2, 0, 0.3) ingots are produced by arc melting high purity (at least 99.9%) iron blocks (4 N) and tin pellets (5 N) under argon protection. To obtain better chemical homogeneity, each ingot is melted at least 4 times. Subsequently, each sample is remelted in a quartz tube in an argon atmosphere and ribbons are obtained using a melt-spinning technique. The surface velocity of the copper wheel is 12.5 m/s and the ribbon thickness is about 60 μm. The iron-tin phase diagram [21,26,28,30,31] shows that $Fe_3Sn$ only exists within a very narrow temperature window between 1023 ~ 1113 K, $Fe_5Sn_3$ will appear above 1086 K, and $Fe_3Sn_2$ will appear below 1043 K. Our ribbons are vacuum treated and annealed at

800 °C for 5 days, and then quenched into ice water. The crystal structure is characterized by X-ray diffraction (XRD), using Cu $K_\alpha$ radiation ($\lambda$ =1.5418 Å). The comparative information of the composition distribution of the sample is obtained by backscattered electron (BSE) and elemental analysis is done using energy dispersive spectroscopy (EDS). Magnetic and magnetotransport measurements are carried out on a 9 T physical properties measurement system (Quantum Design DynaCool), using the standard four-probe methods. To eliminate the longitudinal voltage contribution due to any misalignment of the connections, the Hall resistivity is measured by applying both positive and negative magnetic fields, and the average Hall resistivity is calculated by $\rho_H = \frac{\rho_H(H) - \rho_H(-H)}{2}$.

## III. RESULTS AND DISCUSSION

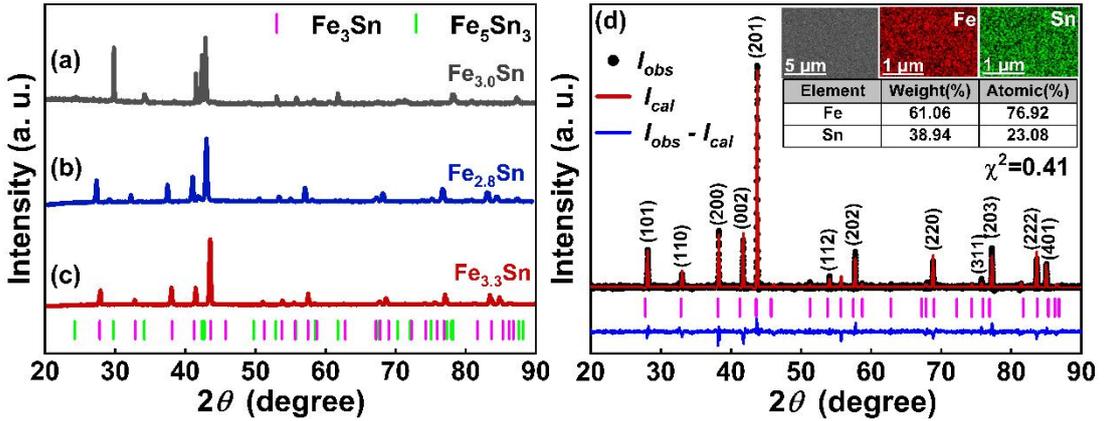

FIG. 1. Crystal structure and composition. XRD patterns of $Fe_{3.0}Sn$ (a) $Fe_{2.8}Sn$ (b) and $Fe_{3.3}Sn$ (c) at room temperature. (d) XRD refinement for $Fe_{3.3}Sn$. The inset shows a backscattered electron (BSE) micrograph in the upper left corner, energy dispersive spectroscopy (EDS) elemental mapping in the upper right corner, and the elemental composition percentage of the $Fe_{3.3}Sn$ sample at the bottom.

Fig.1 (a-c) shows the XRD patterns of $Fe_{3.3}Sn$, $Fe_{2.8}Sn$, and $Fe_{3.0}Sn$ at room temperature. From Fig.1 (a) it can be seen that the stoichiometric $Fe_3Sn$ does not have a single-phase structure, and the percentages of $Fe_5Sn_3$ phase and $Fe_3Sn$ phase are approximately 85% and 15%, respectively. Ref.[27] points out that the average electron/atom ratio $n$ of T in $T_3Sn$ (T represents the transition metal element) determines the stability of the structure. When 7.4 < $n$ < 8 or $n$>9.7, the D0$_{19}$ structure is stable, that is, Fe-rich or Sn-rich conditions are helpful to stabilize the D0$_{19}$ structure. From Fig.1 (b) it can be seen that Fe$_5$Sn$_3$ is greatly suppressed in the Sn-rich Fe$_{2.8}$Sn sample, and its phase ratio is reduced from 85% to 27%. The Fe$_{74.25}$Sn$_{25.75}$ sample in Ref.[23] also shows the same result. In the Fe-rich Fe$_{3.3}$Sn sample shown in Fig.1 (c), Fe$_5$Sn$_3$ can be completely suppressed, thus pure D0$_{19}$ Fe$_3$Sn is obtained, which is consistent with the case of the Fe$_{76.5}$Sn$_{23.5}$ sample by solid phase reaction in Ref.[32]. Therefore, the presence of D0$_{19}$ Fe$_3$Sn is highly dependent on the composition. To further obtain the crystal structure information of Fe$_{3.3}$Sn, Fig.1 (d) shows the refinement results, and the lattice constants are $a = b = 5.4794$ Å, $c = 4.3619$ Å. Both the BSE image and EDS elemental mapping in the inset of Fig.1 (d) show that the Fe$_{3.3}$Sn sample has a homogeneous composition, and its actual composition measured by EDS is Fe$_{76.92}$Sn$_{23.08}$ which is nearly the same as the nominal composition.

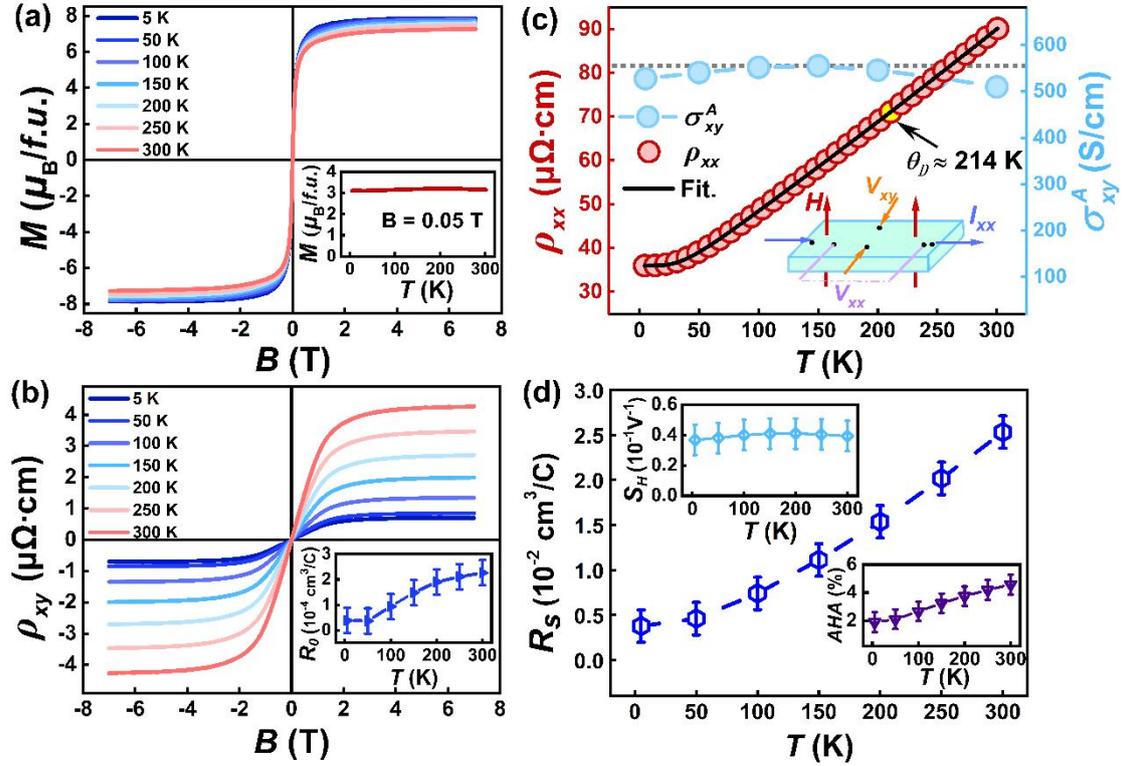

FIG. 2. Magnetic and transport properties of Fe$_{3.3}$Sn. Magnetization $M$ (a) and Hall resistivity $\rho_{xy}$ (b) as a function of magnetic induction $B$ at selected temperatures. The insets in (a) and (b) show

the thermomagnetic curve $M(T)$ under 500 Oe and the temperature $T$ dependence of the ordinary Hall coefficient $R_0$, respectively. (c) Temperature dependence of the longitudinal resistivity $\rho_{xx}$ (left axis) and anomalous Hall conductivity $\sigma_{xy}^A$ (right axis), where $\rho_{xx}(T)$ is fitted using the Bloch-Grüneisen Law (black line) and the grey dashed line represents the theoretically calculated value of the intrinsic anomalous Hall conductivity of polycrystalline Fe$_3$Sn. The inset shows the schematic measurement geometry for $\rho_{xx}$ and $\rho_{xy}$. (d) Temperature dependence of the anomalous Hall coefficient $R_S$, the anomalous Hall scale factor $S_H$ (top inset), and the anomalous Hall angle AHA (bottom inset).

Fig. 2 (a) shows the hysteresis loops and the $M(T)$ curve (inset) from 5 K to 300 K of Fe$_{3.3}$Sn. The external magnetic field is parallel to the ribbon plane to avoid the influence of sample shape on magnetism. The $M(T)$ curve remains at around 3.0 $\mu_B$/f.u. throughout the entire temperature range, indicating that the Fe$_{3.3}$Sn sample does not undergo a magnetic or structural transition, which is the same as Fe$_3$Sn [25]. The $M(H)$ loops below 300 K overlap almost completely, demonstrating that all magnetic parameters are robust to temperature. This result is consistent with previously reported data for polycrystalline and single-crystalline Fe$_3$Sn samples [23,25]. This is due to the high Curie temperature of Fe$_3$Sn, $T_C$ = 743 K [24]. The saturation magnetization $M_S$ of Fe$_{3.3}$Sn obtained from the $M(H)$ loop at 5 K is 7.85 $\mu_B$/f.u., which seems much larger than that of Fe$_3$Sn (6.6 $\mu_B$/f.u.) [25], but its magnetic moment per Fe atom is 2.38 $\mu_B$, which does not conflict with other work.

The schematic measurement geometry for the Hall resistivity $\rho_{xy}$ and the longitudinal resistivity $\rho_{xx}$ is shown in the inset of Fig. 2(c). The current $I$ is applied along the $x$ direction, and the external magnetic field $H$ is perpendicular to the ribbon sample and applied along the $z$ axis. Fig. 2(b) shows the field dependence of the Hall resistivity $\rho_{xy}$ at various temperatures. The schematic measurement geometry is shown in the inset of Fig. 2(c). Fe$_{3.3}$Sn shows an anomalous Hall effect (AHE) with a small ordinary Hall effect (OHE) at each temperature. The total $\rho_{xy}$ presented in Fig. 2(b) can be expressed by the empirical formula $\rho_{xy} = \rho_{xy}^o + \rho_{xy}^A = R_0 B_z + R_S M_z$, where $\rho_{xy}^o$ and $\rho_{xy}^A$ are the ordinary Hall resistivity and the anomalous Hall resistivity, respectively, $R_0$ and $R_S$ are the ordinary Hall coefficient and the anomalous Hall coefficient, respectively, $B_z$ and $M_z$ are the magnetic induction and the magnetization in the z direction, respectively [1,8,9]. Thus, we infer $R_0$ and $\rho_{xy}^A$ from the high field slope (above 4 T) and intercept of $\rho_{xy}$, respectively. The inset of Fig. 2 (b) shows the temperature dependence of $R_0$. As temperature increases, $R_0$ monotonically increases from $0.38 \times 10^{-4}$ cm³/C to $2.26 \times 10^{-4}$ cm³/C. The above temperature behavior of $R_0$ in our polycrystalline $Fe_{3.3}Sn$ sample is neither the same as the temperature-induced sign reversal of $R_0$ in the $Fe_3Sn$ single crystal [22] nor the almost temperature-independent $R_0$ in the $Fe_{2.98}Sn$ single crystal [25]. Further, with the help of $R_0$ one can identify the type of charge carriers $q$ and calculate its density $n$ using the relation $n = \frac{1}{R_0|q|}$. The carrier type in our system is always holes, and $n$ reduces dramatically from $16.4 \times 10^{22}$/cm³ to $2.76 \times 10^{22}$/cm³ with increasing temperature. Considering that $Fe_3Sn$ single crystal is prepared in an Fe-rich environment [22], the composition of single-crystalline $Fe_{2.98}Sn$ is slightly Sn-rich [25], and our polycrystalline $Fe_{3.3}Sn$ sample is Fe-rich, thus we speculate that the type of charge carriers and its density is very sensitive to the composition of $Fe_3Sn$, that is, a small change in the composition can adjust its Fermi surface to be close to the hole pocket or electron pocket of the system.

The extrinsic contribution of AHE depends not only on the impurity density [1,33,34] but also on electron-phonon scattering [10,25,33,35-40]. Thus, the temperature $T$ dependence of the longitudinal resistivity $\rho_{xx}$ is measured, as shown in the left axis of Fig. 2 (c). The residual resistivity $\rho_0$ of the polycrystalline $Fe_{3.3}Sn$ ribbon at 5 K is about 35 μΩ cm, which is similar to those of the single crystal samples [22,25] and polycrystalline ingot [23], indicating that the impurity density of these samples is almost the same. The Bloch-Grüneisen Law [41-43] $\rho_{xx} = \rho_0 + \beta\left(\frac{T}{\theta_D}\right)^n \int_0^{\theta_D/T} \frac{x^n dx}{(e^x-1)(1-e^{-x})}$ is used to fit the $\rho_{xx}(T)$ curve and describe the influence of the interaction between electrons and phonons on resistivity, where $\rho_0$ is the residual resistivity, $\beta$ is the electron-phonon coupling constant, $\theta_D$ is the Debye temperature which is the characteristic temperature of electron-phonon scattering in the system. The black solid line in Fig. 2 (c) is the fitting result, and the obtained parameters are $\rho_0 = 35.9$ μΩ cm, $\beta = 144.8$ μΩ cm, $\theta_D = 214$ K, $n = 4.85$. Therefore, electron-phonon scattering can effectively suppress the extrinsic contribution of AHE, when the temperature is higher than 214 K, while the extrinsic contribution will not be washed out by electron-phonon scattering at temperatures below 214 K.

The anomalous Hall conductivity $\sigma_{xy}^A$ is calculated using the formula $\sigma_{xy}^A = -\frac{\rho_{xy}^A}{\rho_{xx}^2 + \rho_{xy}^2} \approx -\frac{\rho_{xy}^A}{\rho_{xx}^2}$. The right axis of Fig. 2(c) shows the temperature dependence of $\sigma_{xy}^A$. The $\sigma_{xy}^A(T)$ curve of our sample is significantly different from other works in the following two aspects. On the one hand, we notice that $\sigma_{xy}^A$ of our sample is almost independent of temperature over the entire temperature range, unlike the sudden drop of $\sigma_{xy}^A$ in Sn-rich Fe$_3$Sn below 200 K. With decreasing temperature from 200 K, the value of $\sigma_{xy}^A$ decreases sharply from 420 S/cm to 265 S/cm in Fe$_{2.98}$Sn single crystal [25] and from 500 S/cm to 200 S/cm in polycrystalline ingot Fe$_{2.88}$Sn [23]. On the other hand, we observe a large $\sigma_{xy}^A$. The maximum value of $\sigma_{xy}^A$ can reach 555 S/cm, and it can still reach 528 S/cm even at 5 K. All these values are very close to the intrinsic value 555 S/cm calculated for polycrystalline Fe$_3$Sn in Ref.[23], as shown by the grey dashed line in the Fig2. (c). Both the temperature behavior of $\sigma_{xy}^A$ and the vaue of $\sigma_{xy}^A$ suggest that the AHE is dominated by the intrinsic contribution in our sample.

To further describe the AHE of our system, the anomalous Hall scaling factor $S_H = \frac{\mu_0 R_s}{\rho_{xx}^2}$ and the anomalous Hall angle $\text{AHA} = \frac{\sigma_{xy}^A}{\sigma_{xx}} \times 100(\%)$ are calculated. Fig. 2 (d) shows the anomalous Hall coefficient $R_S$ plotted as a function of temperature. The top inset in Fig. 2(d) presents the temperature dependence of $S_H$. We observe that $S_H$ is almost temperature-independent within the error bars. For the AHE dominated by the intrinsic mechanism, $S_H$ should be constant and independent of temperature

[11,44,45]. This again suggests that the intrinsic contribution should be dominant in our system. Moreover, the value of $S_H = 0.04 \pm 0.01$ V$^{-1}$ is slightly larger than that of Fe$_{2.98}$Sn single crystal ($0.03 \pm 0.01$ V$^{-1}$) [25] . The bottom inset in Fig. 2 (d) shows the temperature-dependent AHA. We notice that the low-temperature AHA is ∼ 2%, which is also slightly larger than that of other Fe$_3$Sn samples, including Fe$_3$Sn ($\approx$ 1.2%) [22], Fe$_{2.98}$Sn ($\approx$ 0.9%) [25], Fe$_{2.88}$Sn ($\approx$ 0.4%) [23].

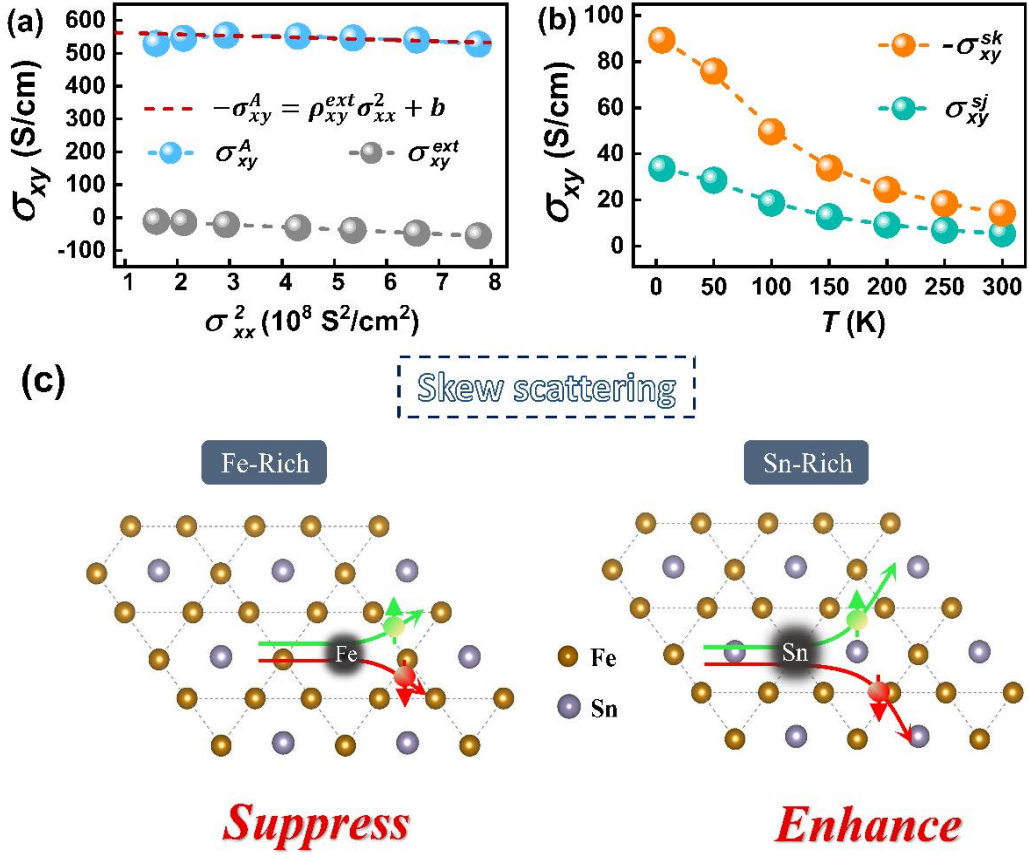

FIG. 3. Anomalous Hall transport analysis of Fe$_{3.3}$Sn. (a) $\sigma_{xx}^2$ dependence of the anomalous Hall conductivity $\sigma_{xy}^A$ and the extrinsic anomalous Hall conductivity $\sigma_{xy}^{ext}$. The dashed line is a linear fitting using the relation $\sigma_{xy}^A = \rho_{xy}^{ext}\sigma_{xx}^2 + b$. (b) Skew-scattering $\sigma_{xy}^{sk}$ and side-jump $\sigma_{xy}^{sj}$ contributions extracted from $\sigma_{xy}^A$ as functions of temperature. (c) Schematic diagram of the effect of different impurity centers on the skew scattering in the kagome magnet Fe$_3$Sn.

The TYJ model $-\sigma_{xy}^A = \rho_{xy}^{ext}\sigma_{xx}^2 + b$ [10,37] is employed to separate the intrinsic and extrinsic contributions from the total anomalous Hall conductivity $\sigma_{xy}^A$, where

$\rho_{xy}^{ext}\sigma_{xx}^2$ represents the extrinsic part and $b$ is the intrinsic part which should be independent of temperature. The blue solid spheres and the red dashed line in Fig. 3 (a) are the experimental data of $\sigma_{xy}^A$ and its fitting results, respectively, and $\rho_{xy}^{ext} =$ -0.026 μΩ cm and $b$ = 566 S/cm. We notice that the intrinsic value $\sigma_{xy}^{in}$ = 566 S/cm obtained by experimental fitting is slightly larger than that obtained by theoretical calculation for stoichiometric Fe$_3$Sn (555 S/cm) [23]. We speculate that the increase in $\sigma_{xy}^{in}$ may be related with the tune of the Fermi level brought by the electron doping in our Fe$_{3.3}$Sn sample. The extrinsic anomalous Hall conductivity $\sigma_{xy}^{ext} = \rho_{xy}^{ext}\sigma_{xx}^2$ is also obtained by the fitting result as shown in Fig. 3 (a). It can be seen that $\sigma_{xy}^{ext}$ is much smaller compared with $\sigma_{xy}^{in}$ and its maximum value does not exceed 50 S/cm, demonstrating that the intrinsic mechanism is dominant and the extrinsic contribution has been almost completely suppressed in our Fe$_{3.3}$Sn sample. We also noticed that $\sigma_{xy}^{ext}$ in the Sn-rich Fe$_3$Sn single crystal is as high as 250 S/cm [25], while $\sigma_{xy}^{ext}$ in the Fe-rich Fe$_3$Sn single crystal is ~ 100 S/cm [46]. Therefore, the Fe-rich composition can effectively inhibit the extrinsic contribution.

Based on the method provided in references [6,9], we further separated the side jump $\sigma_{xy}^{sj}$ and the skew scattering $\sigma_{xy}^{sk}$ contributions of $\sigma_{xy}^{ext}$, as shown in Fig. 3 (b). It is found that $\sigma_{xy}^{sk}$ and $\sigma_{xy}^{sj}$ are of opposite and the same signs as $\sigma_{xy}^{in}$, respectively, that is, the skew scattering contribution would suppress the total AHE. As the temperature increases, both $|\sigma_{xy}^{sk}|$ and $|\sigma_{xy}^{sj}|$ decrease sharply and approach a constant value around $\theta_D = 217$ K, indicating that the electron-phonon scattering also strongly suppresses the extrinsic contribution in our sample as in other works [22,23,25]. Interestingly, due to this electron-phonon scattering suppression, $\sigma_{xy}^A$ showed a sharp drop at low temperatures in other works on Fe$_3$Sn [22,23,25], but this phenomenon was not found in our work. What is the reason? Because the intrinsic contribution does not change with temperature, the above different phenomena should come from the extrinsic contribution. Fig. 3 (b) shows that the extrinsic contribution is mainly provided by the skew scattering. Therefore, analyzing the skew scattering contribution in different works is key. The skew scattering are event occurring at an impurity site, thus the impurity center is crucial for the skew scattering [47]. Fig. 3 (c) shows the schematic diagram of the effect of different impurity centers on the skew scattering in the kagome magnet Fe$_3$Sn, where the yellow and purple spheres represent Fe and Sn atoms occupying the normal sites, respectively, and the small and large dark grey spheres represent the impurity centers of Fe and Sn, which occupy the Sn and Fe sites in the Fe-rich and Sn-rich Fe$_3$Sn systems, respectively.

The anomalous Hall conductivity from skew scattering $\sigma_{xy}^{sk}$ [7-9,34,48] can be expressed as:

$$\sigma_{xy}^{sk} = S\sigma_{xx} \quad (1)$$

$$S \sim \frac{\varepsilon_{SO} v_{imp}}{W^2} \quad (2)$$

$$\varepsilon_{SO} \approx -\frac{\mu_0 \mu_B^2 Z^4}{4\pi a_0^3} \quad (3)$$

where $S$ is the skewness factor, $\varepsilon_{SO}$ is the spin-orbit interaction energy, $W$ is the bandwidth, $v_{imp}$ is the impurity potential strength, Z is the atomic number, and a$_0$ is the Bohr radius [49-51]. The above theory on $\sigma_{xy}^{sk}$ is used here. For the Fe$_3$Sn system, the $\sigma_{xx}$ value in our Fe-rich sample is nearly equal to that in the Sn-rich samples [22,23,25], thus $\sigma_{xy}^{sk}$ is mainly determined by $S$ based on formula (1). Considering that whether it is Fe-rich or Sn-rich Fe$_3$Sn, their $W$ is basically the same in the formula (2), it can be inferred that $\sigma_{xy}^{sk}$ is dominated by $\varepsilon_{SO}$. The $\varepsilon_{so}$ provided by Sn as an impurity center is 14 times that of Fe as an impurity center according to the formula (3). As a result, the conduction electrons will be deflected much more after collision on the Sn impurity center than the Fe impurity center, as shown in Fig. 3 (c).

Therefore, the skew scattering contribution in the Fe-rich sample is suppressed, while that in the Sn-rich Fe₃Sn is enhanced.

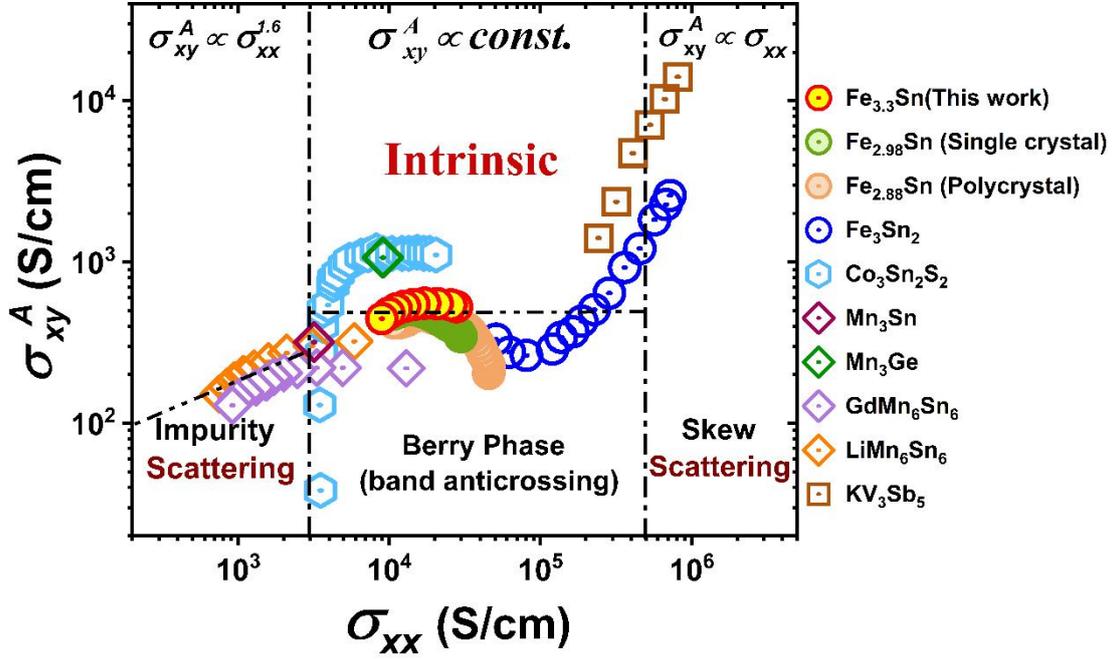

FIG. 4. Longitudinal conductivity $\sigma_{xx}$ dependence of anomalous Hall conductivity $\sigma_{xy}^A$ for kagome magnet in the framework of the unified theory [34,48,52-56]. Here we use the absolute amplitudes of $\sigma_{xy}^A$ without the negative sign. The data of other materials are taken from references and therein [13-19,23,25,57].

To compare the AHE of Fe-rich Fe₃Sn in this work with that of other kagome magnets, $\sigma_{xy}^A$ versus $\sigma_{xx}$ for various materials spanning the three regimes from the dirty regime through to the skew scattering regime are plotted in Fig. 4. The Fe₃Sn system, including polycrystalline Fe$_{3.3}$Sn (this work), single crystalline Fe$_{2.98}$Sn [25], and polycrystalline Fe$_{2.88}$Sn [23], are all in the intrinsic regime. In this regime, $\sigma_{xy}^A$= constant is dominated by Berry phase curvature, not by the extrinsic scattering events according to the unified theory [34,48,52-56]. Obviously, the AHE in the Fe-rich sample satisfies the unified theory, while the AHE in the Sn-rich sample still deviates from the unified theory even though it is in the intrinsic regime. From Fig.4, we can also see that the AHE of Fe₃Sn₂ [19] and KV₃Sb₅ [57] also deviate from the unified theory. We speculate that this may be because Sn and Sb as impurity centers provide strong spin-orbit interaction energy, so the impurity center will affect the AHE of the system. Therefore, researchers engaged in topological device research should appropriately regulate the impurity centers.

## IV. CONCLUSION

In summary, we prepared Fe$_{3\pm x}$Sn polycrystalline samples and found that Fe-rich Fe$_3$Sn is more conducive to stabilizing the hexagonal structure. The magnetic and transport properties of the Fe-rich samples were measured. The experimental results show that its saturation magnetization can reach 7.85 $\mu_B$/*f.u.* at 5 K, and the anomalous Hall conductivity (AHC) shows a temperature-independent characteristic. The maximum value of AHC reaches 555 S/cm, equal to the theoretically calculated value. After fitting AHC, the intrinsic anomalous Hall conductivity $\sigma_{xy}^{in}$ is 566 S/cm, and the extrinsic anomalous Hall conductivity $\sigma_{xy}^{ext}$ is only 50 S/cm. We fitted the longitudinal resistivity using the Bloch-Grüneisen Law and found that the Debye temperature $\theta_D \sim$ 217 K, below which $\sigma_{xy}^{ext}$ will not be washed out by electron-phonon scattering. Further analysis of $\sigma_{xy}^{ext}$ revealed that the skew scattering contribution $\sigma_{xy}^{sk}$ compensated for $\sigma_{xy}^{in}$. We use formulas (1), (2) and (3) to conduct a theoretical analysis of $\sigma_{xy}^{sk}$ and find that the key to regulating $\sigma_{xy}^{sk}$ is to change the impurity center. In addition, we speculate from the data in Fig.4 that impurity centers with strong spin-orbit interactions can cause the system to deviate from the unified theory.

## REFERENCES:

[1]  N. Nagaosa, J. Sinova, S. Onoda, A. H. MacDonald, and N. P. Ong, Reviews of Modern Physics **82**, 1539 (2010).
[2]  D. Xiao, M.-C. Chang, and Q. Niu, Reviews of Modern Physics **82**, 1959 (2010).
[3]  S. Nakatsuji, AAPPS Bulletin **32**, 32, 25 (2022).
[4]  N. Chowdhury, K. I. A. Khan, H. Bangar, P. Gupta, R. S. Yadav, R. Agarwal, A. Kumar, and P. K. Muduli, Proceedings of the National Academy of Sciences, India Section A: Physical Sciences **93**, 477 (2023).

[5] J. P. Lloyd and G. E. Pake, Physical Review **92**, 1576 (1953).
[6] R. Karplus and J. M. Luttinger, Physical Review **95**, 1154 (1954).
[7] J. M. LUTTINGER, Physical Review **112**, 738 (1958).
[8] J. SMIT, Physica **21**, 877 (1955).
[9] J. SMIT, Physica **24**, 39 (1958).
[10] Y. Tian, L. Ye, and X. Jin, Physical Review Letters **103**, 087206 (2009).
[11] C. Zeng, Y. Yao, Q. Niu, and H. H. Weitering, Physical Review Letters **96**, 037204 (2006).
[12] J.-X. Yin, B. Lian, and M. Z. Hasan, Nature **612**, 647 (2022).
[13] S. Nakatsuji, N. Kiyohara, and T. Higo, Nature **527**, 212 (2015).
[14] A. K. Nayak *et al.*, Science Advances **2**, e1501870 (2016).
[15] C. L. Dong Chen, Chenguang Fu, Haicheng Lin, Walter Schnelle, Yan Sun, and Claudia Felser, Physical Review B **103**, 144410 (2021).
[16] T. Asaba, S. M. Thomas, M. Curtis, J. D. Thompson, E. D. Bauer, and F. Ronning, Physical Review B **101**, 174415 (2020).
[17] E. Liu *et al.*, Nature Physics **14**, 1125 (2018).
[18] S. S. Qi Wang, Xiao Zhang, Fei Pang and Hechang Lei, Physical Review B **94**, 075135 (2016).
[19] L. Ye *et al.*, Nature **555**, 638 (2018).
[20] Y. W. Zheng Li, Appl. Phys. Lett **122**, 032401 (2023).
[21] C. Shen, I. Samathrakis, K. Hu, H. K. Singh, N. Fortunato, H. Liu, O. Gutfleisch, and H. Zhang, npj Computational Materials **8**, 248, 248 (2022).
[22] B. P. Belbase, L. Ye, B. Karki, J. I. Facio, J.-S. You, J. G. Checkelsky, J. van den Brink, and M. P. Ghimire, Physical Review B **108**, 075164 (2023).
[23] T. S. Chen *et al.*, Science Advances **8**, eabk1480, eabk1480 (2022).
[24] P. P. J. V. E. a. K. H. J. BUSCHOW, Journal of the Less-Common Metals **159**, 1 (1990).
[25] A. Low, S. Ghosh, S. Ghorai, and S. Thirupathaiah, Physical Review B **108**, 094404, 094404 (2023).
[26] L. D. K.C. Hari Kumar. P. Wollants, Calphad **20**, 139 (1996).
[27] K. Kanematsu, Transactions of the Japan Institute of Metals **27**, 225 (1986).
[28] B. Fayyazi, K. P. Skokov, T. Faske, D. Y. Karpenkov, W. Donner, and O. Gutfleisch, Acta Materialia **141**, 434 (2017).
[29] G. T. a. E. Both, Physical Review B **2**, 3477 (1970).
[30] S. B. M. SINGH, Journal of Materials Science Letters **5**, 733 (1986).
[31] H. Giefers and M. Nicol, Journal of Alloys and Compounds **422**, 132 (2006).
[32] C. Echevarria-Bonet, N. Iglesias, J. S. Garitaonandia, D. Salazar, G. C. Hadjipanayis, and J. M. Barandiaran, Journal of Alloys and Compounds **769**, 843 (2018).
[33] A. Shitade and N. Nagaosa, Journal of the Physical Society of Japan **81**, 083704, 083704 (2012).
[34] S. Onoda, N. Sugimoto, and N. Nagaosa, Physical Review Letters **97**, 126602 (2006).
[35] A. Bid, A. Bora, and A. K. Raychaudhuri, Physical Review B **74**, 035426 (2006).
[36] R. Sharma, M. Bagchi, Y. Wang, Y. Ando, and T. Lorenz, Physical Review B **109**, 104304 (2024).
[37] D. Hou, G. Su, Y. Tian, X. Jin, S. A. Yang, and Q. Niu, Physical Review Letters **114**, 217203 (2015).

[38] J. Qi *et al.*, Physical Review B **104**, 214417 (2021).
[39] D. Ködderitzsch, K. Chadova, J. Minár, and H. Ebert, New Journal of Physics **15** (2013).
[40] S. Y. Liu, N. J. M. Horing, and X. L. Lei, Physical Review B **76**, 195309 (2007).
[41] J. M. Ziman, *Electrons and Phonons* (Clarendon Press, Oxford, 1960).
[42] J. Callaway, *Quantum Theory of the Solid State* (Elsevier, Amsterdam, 1991).
[43] M. Tsuji, Journal of the Physical Society of Japan **12**, 828 (1957).
[44] N. Manyala, Y. Sidis, J. F. Ditusa, G. Aeppli, D. P. Young, and Z. Fisk, Nature Materials **3**, 255 (2004).
[45] S. S. Qi Wang Physical Review B **94**, 075135 (2016).
[46] L. Prodan, A. Chmeruk, L. Chioncel, V. Tsurkan, and I. Kézsmárki, Physical Review B **110**, 094407 (2024).
[47] Y. Niimi and Y. Otani, Reports on Progress in Physics **78** (2015).
[48] S. Onoda, N. Sugimoto, and N. Nagaosa, Physical Review B **77**, 165103 (2008).
[49] J. M. D. COEY, *Magnetism and Magnetic Materials* (Cambridge University Press, 2009).
[50] H. S. J. Stohr, *Magnetism From Fundamentals to Nanoscale Dynamics* (Springer, 2006).
[51] W. J. Fan, L. Ma, and S. M. Zhou, Journal of Physics D: Applied Physics **48**, 195004 (2015).
[52] T. Miyasato, N. Abe, T. Fujii, A. Asamitsu, S. Onoda, Y. Onose, N. Nagaosa, and Y. Tokura, Physical Review Letters **99**, 086602 (2007).
[53] Z. A. Nishchhal Verma, Mohit Randeria, Science Advances **8**, eabq2765 (2022).
[54] A. A. Burkov and L. Balents, Physical Review Letters **91**, 057202 (2003).
[55] P. Mitra, R. Misra, A. F. Hebard, K. A. Muttalib, and P. Wölfle, Physical Review Letters **99**, 046804 (2007).
[56] N. S. Shigeki Onoda, Naoto Nagaosa, Progress of Theoretical Physics **116**, 61 (2006).
[57] S.-Y. Yang *et al.*, Science Advances **6**, eabb6003 (2020).